# INTERNAL ROTATIONS AND STRESS TENSOR SYMMETRY IN THEORIES OF NEMATIC LIQUIDS AND SOLIDS

## A. I. Leonov[a1] and V.S. Volkov[b]

a) *Department of Polymer Engineering, The University of Akron,*
*Akron, Ohio 44325-0301, USA.*

b) *Laboratory of Rheology, Institute of Petrochemical Synthesis, Russian*
*Academy of Sciences, Leninsky Pr., 29, Moscow 117912 Russia.*

**Abstract**

The paper analyses kinematics and dynamics of internal rotations with spin and their effects on the constitutive relations for uniaxial (nematic) liquids and for weakly elastic nematic solids. It is shown that neglecting the internal rotational inertia terms and effects of director gradient made the stress symmetric. This not only highly simplifies the theories but also allows calculating all the kinematic variables of internal rotations without any additional constants, other than those presented in the simplified theory with symmetric stress.

## Introduction.

An important feature of liquids and solids with uniaxial rigid particles (molecules for molecular media or macroscopic particles for suspensions and composites) is that the particles have an additional degree of freedom, their internal rotations. The problem on how to incorporate this effect in continuum approach was long a subject of studies.

Many aspects of the problem were elaborated for liquids. The constitutive equation for isotropic fluids with spherical particles having internal angular momentum was first proposed by Born [1] and rediscovered later by Sorokin [2]. Grad [3] was the first who applied the methods of irreversible thermodynamics to analysis of such fluids. Then de Groot and Mazur extensively discussed the problem in their text [4]. Ericksen [5] was the first who developed the theory of uniaxial liquids with an anisotropic viscosity and yield, caused by orientation of the medium at the rest state. He used in his theory the symmetric stress and a macroscopic unit vector called director, which described internal rotations of uniaxial particles in liquid. In the following the reference to the model of Ericksen liquid will mean his constitutive model without yield. One can also find an extensive review of other Ericksen papers related to the topic in the text by Truesdell and Noll [6] (Sect. 127). Allen and de Silva [7] extended the Ericksen approach. They analysed the macroscopic effects of rotating uniaxial particles including the internal rotational inertia of particles and internal couples, which may cause non-symmetry of stress. Leslie [8] rationalized and simplified the theory [7] in what is known today the Leslie-Ericksen model. In this model the stress is generally non-symmetric. A detailed thermodynamic derivation of constitutive equations for flows of low molecular nematics, made in de Gennes and Prost text [9] (Ch.5), resulted in Leslie constitutive equations [8] with non-symmetric stress. The text [9] demonstrated that for the nematic liquid crystals, the only contribution in internal couples is effect of space gradient of the director. It was noted in several publications [10-12] that the equilibrium stress in nematic liquid crystals is defined ambiguously. Therefore it was also proved (e.g. see [12]) that there exists a unique way to define the equilibrium stress tensor in the system in such a way that it is symmetric. In spite of its rigor, we will not use in the following this approach, and consider in accordance with [9] the equilibrium stress in nematic liquid crystals as generally non-symmetric. Some effects of internal rotations in the viscoelastic "uniaxial liquids" (i.e. those with uniaxially symmetric particles) were recently considered in paper [13].

---

[1] Corresponding author. Fax: +011-330-258-2339. *E-mail:* leonov@uakron.edu (A.I. Leonov)

De Gennes [14] analyzed first the behavior of liquid crystalline crosslinked elastomers as weakly elastic nematic solids, including into analysis the director gradient. In many publications after that Warner (e.g. see the last paper [15] and references there) has developed a theory with finite strains, and finite internal and frame (body) rotations, omitting however, the terms of the director space gradient. Remarkably, this approach leads to a good comparison of theory with experiments for liquid crystalline elastomers [15].

The present paper analyses the kinematics and dynamics of internal rotations and their effects on the constitutive relations for nematic liquids with an anisotropic viscosity, and for weakly elastic nematic solids. Sections 3 and 4 demonstrate that neglecting the internal rotational inertia and effects of director gradient not only highly simplifies the theory, making the stress symmetric, but also allows calculating all the kinematic variables of internal rotations without any additional constants, other than those presented in the simplified theory with symmetric stress.

## 2. Internal rotations in uniaxial continuum

### 1. Kinematics of internal rotations

All the continuum theories assume that it is possible to describe macroscopically the effects of orientation and internal rotations of particles as evolution of the unit vector $\underline{n}$ called director, and its rigid rotations [7] with angular velocity $\underline{\omega}^I$. It is convenient to decompose the angular velocity $\underline{\omega}^I$ in the sum of two vectors, the spin $\underline{\omega}^I_\parallel$ and an additional vector $\underline{\omega}^I_\perp$, i.e. $\underline{\omega}^I = \underline{\omega}^I_\parallel + \underline{\omega}^I_\perp$. The spin is defined as directed along the director being represented as $\underline{\omega}^I_\parallel = \omega^I_\parallel \underline{n}$. It characterises in average the angular velocity of revolution of particles around their axis. The (scalar) angular velocity of spin $\omega^I_\parallel$ is then defined as the projection of the total internal angular velocity on the direction of director, i.e. $\omega^I_\parallel = \underline{\omega}^I \cdot \underline{n}$. Then the additional component $\underline{\omega}^I_\perp$ characterises the change in the total internal angular velocity $\underline{\omega}^I$ caused by changing orientation of director. Therefore we call $\underline{\omega}^I$ the "orientational" angular velocity of internal rotations. Evidently, the vectors $\underline{\omega}^I_\parallel$ and $\underline{\omega}^I_\perp$ are orthogonal, i.e.

$$\underline{\omega}^I_\perp \cdot \underline{\omega}^I_\parallel = \omega^I_\parallel \, \underline{n} \cdot \underline{\omega}^I_\perp = \omega^I_\parallel n_k \cdot \omega^I_{\perp k} = 0 \,. \tag{1}$$

We can also split the angular velocity of frame (body) rotations $\underline{\omega}$ in the sum of orthogonal components, $\underline{\omega} = \underline{\omega}_\parallel + \underline{\omega}_\perp$, where $\underline{\omega}_\parallel = \omega_\parallel \underline{n} = (\underline{\omega} \cdot \underline{n})\underline{n}$.

Since the director exercises the rigid rotation, its speed $\underline{\dot{n}}$ is expressed as:

$$\underline{\dot{n}} = \underline{\omega}^I \times \underline{n} = \underline{\omega}^I_\perp \times \underline{n} = -\underline{\underline{\omega}}^I_\perp \cdot \underline{n} \,, \text{ or } \dot{n}_i = \varepsilon_{ijk}\omega^I_{\perp j}n_k = -\omega^I_{\perp ik}n_k \,. \tag{2}$$

Here $\varepsilon_{ijk}$ is the antisymmetric unit tensor and overdot denotes hereafter the time derivative. It is also convenient to introduce the antisymmetric tensors, $\underline{\underline{\omega}}^I, \underline{\underline{\omega}}^I_\parallel$ and $\underline{\underline{\omega}}^I_\perp$, called respectively the total, spin and orientational internal 'vorticities', and also the similar anti-symmetric tensors $\underline{\underline{\omega}}, \underline{\underline{\omega}}_\parallel$ and $\underline{\underline{\omega}}_\perp$. The components of these tensors are defined as follows:

$$\omega_{ik}^I \equiv \varepsilon_{ikj}\omega_j^I, \quad \omega_{\|ik}^I \equiv \varepsilon_{ikj}\omega_{\|j}^I, \quad \omega_{\perp ik}^I \equiv \varepsilon_{ikj}\omega_{\perp j}^I; \tag{3}$$

$$\omega_{ik} \equiv \varepsilon_{ikj}\omega_j, \quad \omega_{\|ik} \equiv \varepsilon_{ikj}\omega_{\|j}, \quad \omega_{\perp ik} \equiv \varepsilon_{ikj}\omega_{\perp j}. \tag{3a}$$

Here the body vorticity $\underline{\underline{\omega}}$ is commonly defined as the anti-symmetric part of the body gradient velocity $\underline{\nabla}\mathbf{v}$ being split into symmetric, $\underline{\underline{e}}$ and anti-symmetric $\underline{\underline{\omega}}$ parts. The equalities, $\underline{\underline{\omega}}^I = \underline{\underline{\omega}}_\|^I + \underline{\underline{\omega}}_\perp^I$, and $\underline{\underline{\omega}} = \underline{\underline{\omega}}_\| + \underline{\underline{\omega}}_\perp$ hold due to Eqs.(3) and (3a) and the additivity of the vector components of body and internal angular velocities. According to Eqs.(3) and (3a) and the definitions of respective orthogonal vector components for total and internal angular velocities,

$$\underline{\underline{\omega}}_\|^I \cdot \underline{n} = \underline{\underline{\omega}}_\| \cdot \underline{n} = 0, \text{ therefore } \underline{\underline{\omega}}^I \cdot \underline{n} = \underline{\underline{\omega}}_\perp^I \cdot \underline{n}, \quad \underline{\underline{\omega}} \cdot \underline{n} = \underline{\underline{\omega}}_\perp \cdot \underline{n}. \tag{4}$$

We now define the relative vorticity $\underline{\underline{\omega}}^r$ in the uniaxially symmetric anisotropic media as

$$\underline{\underline{\omega}}^r \equiv \underline{\underline{\omega}} - \underline{\underline{\omega}}^I = \underline{\underline{\omega}}_\|^r + \underline{\underline{\omega}}_\perp^r. \tag{5}$$

Note that the relative vorticity $\underline{\underline{\omega}}^r$ is frame invariant. Multiplying Eq.(5) scalarly by $\underline{n}$ and using Eq. (4) yields:

$$\underline{\underline{\omega}}^r \cdot \underline{n} = \underline{\underline{\omega}}_\perp^r \cdot \underline{n} = \overset{0}{\underline{n}}, \quad \overset{0}{\underline{n}} = \dot{\underline{n}} - \underline{n} \cdot \underline{\underline{\omega}}. \tag{6}$$

Here $\overset{0}{\underline{n}}$ is the Jaumann derivative of director $\underline{n}$.

Eqs. (2) and (3) also allow expressing $\underline{\underline{\omega}}_\perp^I$ and $\underline{\underline{\omega}}_\perp^I$ via $\underline{n}$ and $\dot{\underline{n}}$ as:

$$\omega_{\perp i}^I = \varepsilon_{ijk}n_j\dot{n}_k, \quad \underline{\underline{\omega}}_\perp^I = \underline{n}\dot{\underline{n}} - \dot{\underline{n}}\underline{n} \quad (\left|\underline{\underline{\omega}}_\perp^I\right| = \left|\dot{\underline{n}}\right|). \tag{7}$$

Similar expressions for the $\underline{\underline{\omega}}_\perp^r$ and $\underline{\underline{\omega}}_\perp^r$ via $\underline{n}$ and $\overset{0}{\underline{n}}$ following from Eqs.(3) and (6) are:

$$\omega_{\perp i}^r = \varepsilon_{ijk}\overset{0}{n}_j n_k, \quad \underline{\underline{\omega}}_\perp^r = \underline{n}\overset{0}{\underline{n}} - \overset{0}{\underline{n}}\underline{n}. \tag{7a}$$

Formulae (7) and (7a) being independent of physical models of nematic media play important role in various physical cases. Two of them are demonstrated below.

It should be mentioned that for elastic solids of nematic type, additional geometric description is needed, which combine finite strains with finite total and internal rotations. For small elastic deformations and rotations, such a description is developed in Section 4.

## 2.2. Balance laws and dissipation

The continuum theories of media with internal rotations are based on the balance laws of linear and angular momentum [7]:

$$\rho \dot{\underline{v}} = \underline{\nabla} \cdot \underline{\underline{\sigma}} \tag{8a}$$

$$\rho \dot{\underline{\underline{S}}} = 2\underline{\underline{\sigma}}^a + \underline{\underline{m}}, \quad S_{ij} = \varepsilon_{ijk} L_k \ , \tag{8b}$$

where

$$m_{ij} = \varepsilon_{ijk} \partial \mu_{ek} / \partial x_e . \tag{9}$$

Here $\rho$ is the mass density, $\underline{v}$ is the velocity, $\underline{\nabla} \cdot \underline{\underline{\sigma}}$ is the divergence of stress tensor, $\underline{\underline{S}}$ is the internal moment of momentum, $\underline{L}$ is the internal angular moment, and $\underline{\underline{\sigma}}^a$ is the anti-symmetric part of the stress tensor. Here we did not include into analysis the thermal and electromagnetic effects. The stressed state of a continuum with internal rotations is characterized by the stress tensor $\underline{\underline{\sigma}}$, and couple stress tensor $\underline{\underline{\mu}}$. Note that while the equation of motion (8a) cannot be in general satisfied when assuming the stress tensor $\underline{\underline{\sigma}}$ being identically equal to zero, the equation of rotational motion (8b) can still be satisfied when neglecting the couple stress tensor $\underline{\underline{\mu}}$. The stress tensors $\underline{\underline{\sigma}}$, and $\underline{\underline{\mu}}$ are generally non-symmetric. Therefore the number of unknowns describing mechanical interactions in a deformed continuum increases from six to eighteen. In the non-polar case, when internal angular momentum, couple stresses, and body moments are zero, the stress tensor according to Eq. (8b) is symmetric.

In isotropic molecular media, the density of internal couples $\underline{\underline{m}}$ can be omitted in Eq.(8b) due to their negligible contribution in the balance equations and in the entropy production [3,4]. These internal couples represent the key issue in the Cosserat non-symmetrical isotropic elasticity [6] (Sect.98), and perhaps because of the results [3], the non-symmetry effects of this theory for molecular isotropic liquids and solids are very small. However, in the nematic case, the internal couples, related to the molecular field, play important role in the equilibrium of low molecular nematic liquid crystals [9].

It is convenient to operate in thermodynamic theories widely accepted in the nematodynamics, the density (per mass unit) of the Helmholtz free energy, which can be approximately represented in the form:

$$f(T, \underline{n}, \underline{\nabla n}, \underline{\underline{\chi}}) \approx f_1(T, \underline{n}, \underline{\nabla n}) + f_2(T, \underline{n}, \underline{\underline{\chi}}) \,. \qquad (10)$$

Here $f_1$ stands for the contribution similar to the low molecular nematic liquids, and $f_2$ for other, e.g. polymeric, contributions to the free energy, characterized by other state variables denoted in general as various tensors of different rank $\underline{\underline{\chi}}$. Evidently, presentation (10) is oversimplified and works properly only in asymptotic cases. Fortunately, these asymptotic cases can be considered for the description of several important physical situations. In accordance with Eq.(10), it is possible to introduce the decomposition of the stress and couple stress into equilibrium and non-equilibrium contributions:

$$\underline{\underline{\sigma}} + p\underline{\underline{\delta}} = \underline{\underline{\sigma}}^e + \underline{\underline{\tilde{\sigma}}}, \qquad \underline{\underline{\mu}} = \underline{\underline{\mu}}^e + \underline{\underline{\tilde{\mu}}} \,. \qquad (11)$$

Here $p$ is the pressure; $\underline{\underline{\sigma}}^e$ (and $\underline{\underline{\tilde{\sigma}}}$) and $\underline{\underline{\mu}}^e$ (and $\underline{\underline{\tilde{\mu}}}$) are respectively the equilibrium and non-equilibrium contributions to the stress and couple stress tensors. Ignoring higher-order effects connected with the angular velocity gradient yields $\underline{\underline{\tilde{\mu}}} = 0$ [7]. This result is similar to that obtained by Leslie [8] for liquid crystals of nematic type. Ericksen [16,17] and Leslie [18] demonstrated that for the case of nematic liquid crystals, the equilibrium contributions to stress and couple stress are:

$$\underline{\underline{\sigma}}^e = -\underline{\underline{\pi}} \cdot \left(\underline{\nabla n}\right)^T, \qquad \underline{\underline{\mu}}^e = -\underline{\underline{\pi}} \cdot \underline{\underline{\varphi}}^T, \qquad (12)$$

where

$$\underline{\underline{\pi}} = \partial f_1 / \partial \underline{\nabla n}, \qquad \underline{\underline{\varphi}} = (\varepsilon_{ijk} n_k) \,.$$

Here $\underline{\nabla} \cdot \underline{n} = \left( \partial n_j / \partial x_i \right)$ is the director gradient, and $^T$ denotes transpose. The effects of the spatial distribution of director orientations are commonly analysed by using the Frank [19] free energy:

$$2 f_1 = K_1 \left( \underline{\nabla} \cdot \underline{n} \right)^2 + K_2 \left( \underline{n} \cdot \underline{\nabla} \times \underline{n} \right)^2 + K_3 \left( \underline{n} \times \underline{\nabla} \times \underline{n} \right)^2 \ . \tag{13}$$

Here $K_i$ being of the dimensionality of force, are the Frank' moduli. The three contributions to $f_1$ in Eq.(13) are associated with the three modes of the distribution of the director gradient: splay, twist, and bend.

As shown in Appendix A, the antisymmetric tensor $\underline{\underline{m}}$ for nematic liquid crystals can be presented in the form:

$$\underline{\underline{m}} = -2 \underline{\underline{\sigma}}^{ea} + \underline{n} \underline{h}^{\perp} - \underline{h}^{\perp} \underline{n} \ . \tag{14}$$

Here $\underline{\underline{\sigma}}^{ea}$ is the anti-symmetric part of the equilibrium stress tensor. The vector $\underline{h}^{\perp}$ is the transverse component of the molecular field. According to de Gennes and Prost [9], the molecular field $\underline{h}$ has the form

$$\underline{h} = \underline{\nabla} \cdot \underline{\underline{\pi}} - \underline{\nabla} f_1 \tag{15}$$

In the continuum theory, it is reasonable to postulate along with the existence of the effective angular velocity $\underline{\omega}^I$, also the existence of a symmetric and positive definite moment-of-inertia density tensor $\underline{\underline{I}}$. Then the internal angular momentum density $\underline{L}$ is commonly defined as:

$$\underline{L} = \underline{\underline{I}} \cdot \underline{\omega}^I \quad ( \underline{\omega}^I \equiv \underline{\underline{I}}^{-1} \cdot \underline{L} ). \tag{16}$$

In the statistical theory [20] these quantities are defined as:

$$\underline{\underline{I}}(\underline{x}, t) = \sum_\alpha \left\langle \underline{\underline{I}}^\alpha \delta_\alpha \right\rangle, \quad \underline{L}(\underline{x}, t) = \sum_\alpha \left\langle \underline{\underline{I}}^\alpha \cdot \underline{\omega}^\alpha \delta_\alpha \right\rangle \tag{16a}$$

Here $\delta_\alpha = \delta(\underline{r}^\alpha - \underline{x})$ is the delta function, and $\underline{r}^\alpha$, $\underline{\underline{I}}^\alpha$ and $\underline{\omega}^\alpha$ are respectively the position, moment of inertia, and rotational velocity of $\alpha$ th molecule. It should be mentioned that the effective angular velocity $\underline{\omega}^I$ is not represented as the ensemble average of $\underline{\omega}^\alpha$, rather it is defined by Eq.(16), even in the statistical approach.

For a uniaxial continuum, where the effective angular velocity $\underline{\omega}^I$ *is defined as the angular velocity of director*, the internal angular momentum is represented as:

$$\underline{L} = \underline{\underline{I}} \cdot \underline{\omega}^I, \quad \underline{\underline{I}} = I_\perp \underline{\underline{\delta}} + (I_\parallel - I_\perp)\underline{n}\underline{n} . \tag{17}$$

Here $I_\perp$ and $I_\parallel$ are the principal values of inertia tensor $\underline{\underline{I}}$; the spin inertia being characterized by $I_\parallel$. Substituting the decomposition $\underline{\omega}^I = \underline{\omega}^I_\parallel + \underline{\omega}^I_\perp$ into the first equation (17) and using the second equation (17) yields:

$$\underline{L} = I_\perp \underline{\omega}^I_\perp + I_\parallel \underline{\omega}^I_\parallel, \qquad \underline{\underline{S}} = I_\perp \underline{\underline{\omega}}^I_\perp + I_\parallel \underline{\underline{\omega}}^I_\parallel . \tag{17a}$$

Direct calculations (see Appendix B) show that

$$\dot{\underline{L}} \cdot \underline{n} = I_\parallel \dot{\omega}^I_\parallel, \quad \dot{\underline{\underline{S}}} \cdot \underline{n} = -\underline{\underline{I}}_\perp \cdot \ddot{\underline{n}} + I_\parallel \omega^I_\parallel \underline{\omega}^I_\perp, \quad \underline{\underline{I}}_\perp = I_\perp(\underline{\underline{\delta}} - \underline{n}\underline{n}) . \tag{18}$$

As follows from the second equation in (18), the vector $\dot{\underline{\underline{S}}} \cdot \underline{n}$ is orthogonal to the vector $\underline{n}$. Multiplying Eq.(8b) by $\underline{n}$ with the use of Eqs.(14) and (18) yields:

$$\rho \left[ \underline{\underline{I}}_\perp \cdot \ddot{\underline{n}} - I_\parallel \omega^I_\parallel \underline{\omega}^I_\perp \right] = \underline{h}^\perp + 2\underline{n}\underline{\underline{\tilde{\sigma}}}^a, \qquad \rho I_\parallel \dot{\omega}^I_\parallel = \varepsilon_{ijk} n_i \tilde{\sigma}^a_{jk} . \tag{19}$$

Kinetic energy per mass unit is now represented as the sum of common translation and internal rotation components. According to Eq.(17) the rotational component $K_r$ of kinetic energy (per mass unit) in the anisotropic system is:

$$2K_r = I_\perp (\omega^I_\perp)^2 + I_\parallel (\omega^I_\parallel)^2 = tr[I_\perp (\underline{\underline{\omega}}^I_\perp)^2 + I_\parallel (\underline{\underline{\omega}}^I_\parallel)^2] . \tag{20}$$

It is shown in Appendix C that the result of differentiating Eq.(20) with respect to time can be represented as:

$$\dot{K}_r = tr(I_\perp \underline{\underline{\omega}}^I_\perp \cdot \underline{\underline{\dot{\omega}}}^I_\perp + I_\parallel \underline{\underline{\omega}}^I_\parallel \cdot \underline{\underline{\dot{\omega}}}^I_\parallel) = tr(\underline{\underline{S}} \cdot \underline{\underline{\omega}}^I) . \qquad (21)$$

Then the balance of kinetic energy of internal rotations is written in the common form [4]:

$$(\rho / 2)\dot{K}_r = (\rho / 2)tr(\underline{\underline{S}} \cdot \underline{\underline{\omega}}^I) = tr(\underline{\underline{\tilde{\sigma}}}^a \cdot \underline{\underline{\omega}}^I) . \qquad (22)$$

We now turn our attention to the irreversible effects. Using Eq.(10) the dissipation in the system is represented as follows:

$$D \equiv TP_S \big|_T = tr(\underline{\underline{\tilde{\sigma}}}^s \cdot \underline{\underline{e}}) - tr(\underline{\underline{\tilde{\sigma}}}^a \cdot \underline{\underline{\omega}}^r) - \rho df_2 / dt \big|_T \quad (\geq 0) , \qquad (23)$$

$$\underline{\underline{\tilde{\sigma}}}^s = \underline{\underline{\sigma}}^s - \underline{\underline{\sigma}}^{es}; \qquad \underline{\underline{\tilde{\sigma}}}^a = \underline{\underline{\sigma}}^a - \underline{\underline{\sigma}}^{ea} . \qquad (24)$$

Hereafter using the arguments presented in [7], we ignore the contribution, $-tr(\underline{\underline{\tilde{\mu}}} \cdot \underline{\underline{v}})$, of irreversible coupled stress $\underline{\underline{\tilde{\mu}}}$ into the dissipation. Here the term $\underline{\underline{v}} = \underline{\nabla}\underline{\underline{\omega}}^I$ is the gradient of angular velocity of internal rotations. In Eqs.(23), (24) $\underline{\underline{\sigma}}^s$ and $\underline{\underline{\sigma}}^{es}$ are the symmetric parts of stress tensor and equilibrium stress tensor, $\underline{\underline{\sigma}}^a$ and $\underline{\underline{\sigma}}^{ea}$ the same for anti-symmetric parts, $\underline{\underline{\omega}}^r$ is the relative vorticity defined in Eq.(5). Thus the quantities $\underline{\underline{\tilde{\sigma}}}^s$ and $\underline{\underline{\tilde{\sigma}}}^a$ represent the non-equilibrium parts of the stress tensor. The detailed derivation of dissipation function for low molecular nematic liquid crystals, when in Eq.(23) $f_2 = 0$ and $\underline{\underline{v}} = 0$, is given in the text [9].

### 3. Liquids with anisotropic viscosity

We now consider the incompressible case when $f_2$ in Eq.(10) depends on $T$ and $\underline{n}$, i.e. $df_2 / dt \big|_T = 0$ in Eq.(23). There exist at least two systems that are described with this specification of $f_2$. The first one where $f_2 = 0$, presents the low molecular LC nematics with the non-equilibrium anisotropic viscous contribution to the stress. In this case one can neglect the molecular inertia terms related to the internal rotations, if

the flow is not too fast (or frequencies of oscillations are not too high). The second case represents the continuum description of suspensions in viscous liquids with suspended uniaxial rigid particles, when the equilibrium molecular terms are absent, i.e. $\tilde{\underline{\underline{\sigma}}}^s = \underline{\underline{\sigma}}^s$ and $\tilde{\underline{\underline{\sigma}}}^a = \underline{\underline{\sigma}}^a$. Hand [18] was the first who investigated the relation between this type of suspensions and Ericksen director theory [5] for transversely isotropic liquids, which ignores any dependence of the stress tensor on the director velocity.

In the incompressible case under study, the symmetric part of non-equilibrium stress tensor $\tilde{\underline{\underline{\sigma}}}^s$ can be considered without loss of generality as deviator. Applying to Eq.(23) the quasi-linear approach of non-equilibrium thermodynamics [4] where the kinetic tensors depend on temperature and director, yields the general quasi-linear constitutive equations:

$$\tilde{\underline{\underline{\sigma}}}^s = a_0\underline{\underline{e}} + a_1[\underline{n}\underline{n}\cdot\underline{\underline{e}} + \underline{\underline{e}}\cdot\underline{n}\underline{n} - 2\underline{n}\underline{n}(\underline{n}\underline{n}:\underline{\underline{e}})] + a_2(\underline{n}\underline{n}\cdot\underline{\underline{\omega}}^r - \underline{\underline{\omega}}^r\cdot\underline{n}\underline{n}) \qquad (25a)$$

$$\tilde{\underline{\underline{\sigma}}}^a = a_2(\underline{n}\underline{n}\cdot\underline{\underline{e}} - \underline{\underline{e}}\cdot\underline{n}\underline{n}) + a_3(\underline{n}\underline{n}\cdot\underline{\underline{\omega}}^r + \underline{\underline{\omega}}^r\cdot\underline{n}\underline{n}) + a_4\underline{\underline{\omega}}^r . \qquad (25b)$$

The form of terms in Eqs.(25a,b) is dictated by the symmetry of the deviator $\tilde{\underline{\underline{\sigma}}}^s$, anti-symmetry of $\tilde{\underline{\underline{\sigma}}}^a$ and $\underline{n} \rightarrow -\underline{n}$ invariance. Here the Onsager symmetry of kinetic coefficients has also been used. The temperature dependent scalar kinetic coefficients $a_k = a_k(T)$ have the dimensionality of viscosity.

Using Eqs.(6) and (7a), one can represent Eqs.(25) in the identical form:

$$\tilde{\underline{\underline{\sigma}}}^s = a_0\underline{\underline{e}} + a_1[\underline{n}\underline{n}\cdot\underline{\underline{e}} + \underline{\underline{e}}\cdot\underline{n}\underline{n} - 2\underline{n}\underline{n}(\underline{n}\underline{n}:\underline{\underline{e}})] - a_2(\overset{0}{\underline{n}}\underline{n} + \underline{n}\overset{0}{\underline{n}}) \qquad (26a)$$

$$\tilde{\underline{\underline{\sigma}}}^a = a_2(\underline{n}\underline{n}\cdot\underline{\underline{e}} - \underline{\underline{e}}\cdot\underline{n}\underline{n}) + (a_3 + a_4)(\overset{0}{\underline{n}}\underline{n} - \underline{n}\overset{0}{\underline{n}}) + a_4\underline{\underline{\omega}}^r_\| . \qquad (26b)$$

Here $\omega^r_{\|ij} = \varepsilon_{ijk}n_k\omega^r$, and $\omega^r = \underline{\omega}\cdot\underline{n} - \omega^l_\|$. Eqs. (26a,b) are slightly different from those used by de Gennes and Frost in incompressible case (see Eqs.5.27 and 5.28 in [9]). First, we included in Eq.(26b) the new Born's [1] term proportional to $a_4$, absent in the de Gennes and Prost equation 5.28. Second, we have used the fact that the symmetric part of stress is deviator. This fact yields the additional relation, $\alpha_1 = -1/2(\alpha_5 + \alpha_6)$, between the Leslie coefficients in the de Gennes and Frost's equation 5.27. Note that constitutive equations (26a,b) are unclosed because the evolution equation for director is still unknown.

Substituting Eq.(25) into Eq.(23) where $df_2/dt\big|_T = 0$, yields the expression for the dissipation as a quadratic form relative to the variables $\underline{e}$ and $\underline{\underline{\omega}}^r$,

$$D = a_0 tr\underline{\underline{e}}^2 + 2a_1 tr[\underline{nn}:\underline{\underline{e}}^2 - (\underline{nn}:\underline{e})^2] + 4a_2 tr[\underline{nn}:(\underline{\underline{\omega}}^r \cdot \underline{e}) - 2a_3 tr[\underline{nn}:(\underline{\underline{\omega}}^r)^2] - a_4 tr(\underline{\underline{\omega}}^r)^2$$
(27)

Here it was taken into account that the trace of square of an anti-symmetric tensor is negative. Simple, but only sufficient, conditions for the dissipation to be positive definite are:

$$a_0, a_1, a_3, a_4 > 0, \quad a_1(a_4 + a_3) > a_2^2;$$
(28)

the parameter $a_2$ being sign indefinite.

Using Eq.(19) along with Eqs.(6) and (26) yields:

$$\rho[\underline{\underline{I}}_\perp \cdot \underline{\ddot{n}} - I_\parallel \omega_\parallel^I \underline{\omega}_\perp^I] = \underline{h}^\perp + \gamma_2[\underline{e} \cdot \underline{n} - \underline{n}(\underline{nn}:\underline{e})] - \gamma_1 \overset{0}{\underline{n}}$$
(29a)

$$\rho I_\parallel \dot{\omega}_\parallel^I = \gamma_3(\underline{\omega} \cdot \underline{n} - \omega_\parallel^I).$$
(29b)

Here $\gamma_1 = 2(a_3 + a_4)$, $\gamma_2 = 2a_2$, $\gamma_3 = 2a_4$ are the rotational viscosities, and $\underline{\omega}$ is the angular velocity vector of the body (frame). Remarkable that Eq.(29b) is the linear equation for the scalar component of the internal spin velocity.

Eqs. (26) and (29) form a closed set of constitutive equations for both the low molecular weight nematic LC's and uniaxial suspensions. Note that Eqs.(29a,b) are convenient for analyzing effects of small amplitude oscillations imposed on an anisotropic rest state with a value of director $n_i^0$.

Neglecting inertia effects of internal rotations in Eqs.(29a,b) simplifies the equations (26a,b) for symmetric and antisymmetric parts of non-equilibrium stress tensor :

$$\underline{\underline{\tilde{\sigma}}}^s = \eta_0 \underline{\underline{e}} + \eta_n [\underline{nn} \cdot \underline{\underline{e}} + \underline{\underline{e}} \cdot \underline{nn} - 2\underline{nn}(\underline{nn} : \underline{\underline{e}})] - \frac{\lambda}{2}(\underline{n} h^\perp + \underline{h}^\perp \underline{n}), \qquad \underline{\underline{\tilde{\sigma}}}^a = (\underline{h}^\perp \underline{n} - \underline{n} h^\perp)/2 \quad (30)$$

$$\overset{0}{\underline{n}} = \frac{1}{\gamma_1} \underline{h}^\perp + \lambda[\underline{\underline{e}} \cdot \underline{n} - \underline{n}(\underline{nn} \cdot \underline{\underline{e}})]. \quad (\lambda = \gamma_2/\gamma_1, \ \eta_0 = a_0, \ \eta_n = a_1 - \lambda a_2) \tag{31}$$

The constitutive equations (30,31) contain the transverse component of the molecular field only. The longitudinal component of $\underline{h}$ has no physical sense. Neglecting effects of director gradient in the second equation in (30) yields the symmetry condition $\underline{\underline{\tilde{\sigma}}}^a = 0$ for the non-equilibrium stress tensor. It is important that in this case the first equation in (30) contains only two independent viscosity coefficients $\eta_0$ and $\eta_n$. This is in a marked contrast to the familiar Ericksen equation [5], which contains three independent viscosities. The first term in the right-hand side of orientation equation (31) describes the relaxation of the director towards its equilibrium value under effect of molecular field. The second term describes the effect of the flow that tends to orient the director. The dimensionless parameter $\lambda$ is close to 1 for the elongated molecules and to −1 for the oblate molecules. In the particular case $\underline{h}_\perp = 0$ Eq.(31) reduces to the Ericksen orientation equation [5].

Neglecting the inertia effect in the equation of rotation motion (29b) and substituting Eq. (31) into Eq. (7a) yields:

$$\underline{\omega}_\parallel^r = 0 \ (\omega_\parallel^l = \omega_\parallel), \ \ \underline{\omega}_\perp^r = \lambda(\underline{\underline{e}} \cdot \underline{nn} - \underline{nn} \cdot \underline{\underline{e}}) + (\underline{h}^\perp \underline{n} - \underline{n}\underline{h}^\perp)/\gamma_1. \tag{32}$$

The first relation in (32) demonstrates that the spin of director coincides with the projection of body angular velocity on the direction of director. This is the general result for non-inertial approximation of internal rotations in nematic liquids.

Substituting Eqs.(30) into Eq.(23a) yields the expression for dissipation:

$$D \equiv tr(\underline{\underline{\tilde{\sigma}}}^s \cdot \underline{\underline{e}} + \underline{h}\overset{0}{\underline{n}}) = \eta_0 tr\underline{\underline{e}}^2 + 2\eta_n tr[\underline{nn} \cdot \underline{\underline{e}}^2 - (\underline{nn} \cdot \underline{\underline{e}})^2] + \underline{h}^{\perp 2}/\gamma_1 \tag{33}$$

Eqs.(32) and (33) have been obtained for two classes of liquids under study when neglecting the inertia effects of internal rotations. Remarkably, the dissipation (33) and the internal rotations (32) are completely determined in this case via the dynamic variables of constitutive equations (30,31) and their constitutive parameters $\eta_0$, $\eta_n$, $\gamma_1$, and $\lambda$. According to Eq.(33) three viscosity coefficients $\eta_0$, $\eta_n$, $\gamma_1$ are positive, and the sign of dimensionless parameter $\lambda$ is indefinite.

## 4. Incompressible weakly elastic nematics

The main objective of this Section is evaluating the effects of negligence of internal couples and the inertia of director in the theory [15] of weakly elastic nematic elastomers.

Consider first the kinematical relation between the initial value of director $\underline{n}_0$ in non-deformed state and its actual value $\underline{n}$ in the deformed state. Since the spin rotation $\underline{\underline{\Omega}}_{\parallel}^I$ of director does not change its orientation, this relation is established via an orthogonal tensor $\underline{\underline{q}}$ as:

$$\underline{n} = \underline{\underline{q}} \cdot \underline{n}_0 ; \qquad \underline{\underline{q}} = \exp(\underline{\underline{\Omega}}_{\perp}^I) . \qquad (34)$$

Here $\underline{\underline{\Omega}}_{\perp}^I$ is the anti-symmetric tensor of finite orientational internal rotations. Although the spin rotation $\underline{\underline{\Omega}}_{\parallel}^I$ of director is absent in the kinematical relation (34), it can play an important role being involved in the free energy formulation.

We consider below the anisotropic weakly elastic case, when the elastic strain tensor $\underline{\underline{E}}$ and the tensors of internal $\underline{\underline{\Omega}}^I$ ($= \underline{\underline{\Omega}}_{\parallel}^I + \underline{\underline{\Omega}}_{\perp}^I$) and total $\underline{\underline{\Omega}}$ rotations are small. In this case, the thermodynamic state variables are the tensors $\underline{\underline{E}}$ and $\underline{\underline{\Omega}}^r$ ($\equiv \underline{\underline{\Omega}} - \underline{\underline{\Omega}}^I$), which are frame invariant. Here the small tensors $\underline{\underline{E}}$ and $\underline{\underline{\Omega}}$ are defined via the displacement vector $\underline{u}$ by the common formulae of linear elasticity:

$$2\underline{\underline{E}} = \nabla \underline{u} + (\nabla \underline{u})^T ; \quad 2\underline{\underline{\Omega}} = \nabla \underline{u} - (\nabla \underline{u})^T . \qquad (35)$$

If the state variables $\underline{\underline{E}}$ and $\underline{\underline{\Omega}}^r$ are known, the total internal and body rotations are known separately and the director in the deformed state is found from Eq.(34) as:

$$\underline{n} \approx (\underline{\underline{\delta}} + \underline{\underline{\Omega}}_{\perp}^I) \cdot \underline{n}_0 . \qquad (36)$$

Following Warner [15], we neglect the dependence of the free energy on the director gradient. It means that in this case $f \approx f_2$ in Eq. (10). Searched within this approach a general relation for the Helmholtz free energy in incompressible anisotropic case should be invariant relative to $\underline{n}_0 \rightarrow -\underline{n}_0$ transformation, quadratic in the state variables, and having vanishing relative rotations in isotropic case. Its general form for the weakly elastic case is:

$$\rho F = \frac{1}{2} G_0 tr \underline{\underline{E}}^2 + \frac{1}{2} G_1 [\underline{n}_0 \underline{n}_0 : \underline{\underline{E}}^2 - (\underline{n}_0 \underline{n}_0 : \underline{\underline{E}})^2]$$
$$- D_1 \underline{n}_0 \underline{n}_0 : (\underline{\underline{\Omega}}^r)^2 + D_2 (\underline{\underline{\Omega}}^r \cdot \underline{\underline{E}} : \underline{n}_0 \underline{n}_0) - \frac{1}{2} D_3 tr (\underline{\underline{\Omega}}^r)^2 \tag{37}$$

Warner [15] has discussed a similar form of free energy in compressible case when the second term in Eq.(37) is split in two independent terms. He missed, however, the first and last terms in Eq.(37). According to Warner [15], the third term ($\sim D_1$) in Eq. (37) is a penalty for internal rotations and the fourth one ($\sim D_2$) reflects the coupling effect. All the coefficients in Eq.(37) have the same energetic scale but they are scaled differently with the scalar order parameter $Q$ as [15]:

$$G_1 \sim G_0, \quad D_1 \sim G_0 Q^2, \quad D_2 \sim G_0 Q \; ; \tag{38}$$

the $Q$-scales for the parameters $G_0$ and $G_3$ being unknown.

If the deformation of a nematic elastic solid is considered far away from the order-disorder phase transition, one can additionally employ the thermodynamic stability conditions, which demand the quadratic form (37) to be positive definite. The sufficient conditions for that are:

$$G_1, G_0, D_1, D_3 > 0, \quad G_1(D_1 + D_3) > D_2^2 \,. \tag{39}$$

Here the parameter $D_2$ is sign indefinite.

One can obtain the stress-strain-relative rotation relations using Eq.(23) for dissipation when demanding the dissipation to vanish in the equilibrium, and using the kinematical formulae: $\underline{\underline{\omega}} = \dot{\underline{\underline{\Omega}}}$ , $\underline{\underline{\omega}}^I = \dot{\underline{\underline{\Omega}}}^I$ . These relations are:

$$\underline{\underline{\sigma}}^{s} = \rho \frac{\partial F}{\partial \underline{\underline{E}}} = G_0 \underline{\underline{E}} + G_1 [\underline{n}_0 \underline{n}_0 \cdot \underline{\underline{E}} + \underline{\underline{E}} \cdot \underline{n}_0 \underline{n}_0 - 2\underline{n}_0 \underline{n}_0 (\underline{\underline{E}} : \underline{n}_0 \underline{n}_0)] + D_2 (\underline{n}_0 \underline{n}_0 \cdot \underline{\underline{\Omega}}^r - \underline{\underline{\Omega}}^r \cdot \underline{n}_0 \underline{n}_0)$$

$$\underline{\underline{\sigma}}^{a} = \rho \frac{\partial F}{\partial \underline{\underline{\Omega}}^{r}} = D_1 (\underline{n}_0 \underline{n}_0 \cdot \underline{\underline{\Omega}}^r + \underline{\underline{\Omega}}^r \cdot \underline{n}_0 \underline{n}_0) + D_2 (\underline{n}_0 \underline{n}_0 \cdot \underline{\underline{E}} - \underline{\underline{E}} \cdot \underline{n}_0 \underline{n}_0) + D_3 \underline{\underline{\Omega}}^r . \qquad (40)$$

As common for the linear theory, the extra stress tensor $\underline{\underline{\sigma}}^{s}$ in (40) is deviator.

Relations (40) along with Eq.(36) form the closed set of constitutive equations. We now simplify these equations neglecting the inertia effects of internal rotations, which makes the stress tensor symmetrical: $\underline{\underline{\sigma}}^{a} = 0$. This case seems to be valid not only for the static but even for dynamic situations, when the frequencies of oscillations are not too high. Since this case is quite similar to that analyzed in the previous Section, we just show below the results of the analysis:

$$\underline{\underline{\sigma}} = \underline{\underline{\sigma}}^{s} = G_0 \underline{\underline{E}} + \tilde{G}_1 [\underline{n}_0 \underline{n}_0 \cdot \underline{\underline{E}} + \underline{\underline{E}} \cdot \underline{n}_0 \underline{n}_0 - 2\underline{n}_0 \underline{n}_0 tr(\underline{\underline{E}} \cdot \underline{n}_0 \underline{n}_0)], \qquad \tilde{G}_1 = G_1 - \Lambda D_2 \quad (>0) \quad (41)$$

$$\rho \tilde{F} = \frac{1}{2} G_0 tr \underline{\underline{E}}^2 + \frac{1}{2} \tilde{G}_1 [tr(\underline{n}_0 \underline{n}_0 \cdot \underline{\underline{E}}^2) - tr^2(\underline{n}_0 \underline{n}_0 \cdot \underline{\underline{E}})] . \qquad (42)$$

$$\Omega^e = n_{0k} \Omega_k , \quad \underline{\underline{\Omega}}^r = \Lambda (\underline{n}_0 \underline{n}_0 \cdot \underline{\underline{E}} - \underline{\underline{E}} \cdot \underline{n}_0 \underline{n}_0) , \qquad \Lambda = D_2 /(D_1 + D_3) \qquad (43)$$

It is easy to show that the symmetric stress tensor (41) can be derived from Eq.(42) using the standard relation, $\underline{\underline{\sigma}} = \rho \partial \tilde{F} / \partial \underline{\underline{E}}$. Eq.(43) along with Eq.(36) determine the director in the deformed state with one sign indefinite parameter $\Lambda$. Algebraic relation (43), where director's spin $\Omega^e$ is equal to the spin component of the body rotation physically means the quick adaptation of a rapid inertial relaxation process described by Eq.(7) to the equilibrium condition. The "spontaneous rotations" found in paper [15] might be explained as the simple example of this adaptation.

Remarkably, Eqs.(41) and (42) do not contain the internal rotations, but uniquely determine the internal rotations by eq.(43) using no additional parameter.

**Appendix A**: Derivation of Eq.(14).

The invariance condition of free energy $f_1$ relative to the rigid rotation results in the following relation [16]:

$$\varepsilon_{ijk}\left( n_j \frac{\partial f_1}{\partial n_k} + n_{j,e} \frac{\partial f_1}{\partial n_{k,e}} + n_{e,j} \frac{\partial f_1}{\partial n_{e,k}} \right) = 0 \tag{A1}$$

Taking into account the expression (12) for the Ericksen stress tensor $\underline{\underline{\sigma}}^e$, Eq. (A1) can be rewritten as follows:

$$\varepsilon_{ijk}(n_j \pi_{ek,e} + n_{j,e}\pi_{ek}) = \varepsilon_{ijk}(n_j h_k^\perp - \sigma_{jk}^e). \tag{A2}$$

Here $h_i^\perp = h_i - n_i n_e h_e$ is the transverse component of the molecular field (15). Using Eqs.(9) and (12) yields:

$$m_{ij} = \varepsilon_{ijk}\varepsilon_{ksn}(n_s \pi_{en,e} + n_{s,e}\pi_{en}) \tag{A3}$$

Substituting (A2) into (A3) results in Eq.(14).

**Appendix B**: Derivation of Eq.(18).

1) Substituting $\underline{\omega}_\parallel^I = \omega_\parallel^I \underline{n}$ into the first equation in (17), differentiated with respect to time, and multiplying scalarly the result by $\underline{n}$, yields:

$$\underline{\dot{L}} \cdot \underline{n} = I_\perp \underline{\dot{\omega}}_\perp^I \cdot \underline{n} + I_\parallel \underline{n} \cdot d/dt(\omega_\parallel^I \cdot \underline{n}) = I_\perp[d/dt(\underline{\omega}_\parallel^I \cdot \underline{n}) - \underline{\omega}_\perp^I \cdot \underline{\dot{n}}] + I_\parallel(\dot{\omega}_\parallel^I + \omega_\parallel^I \underline{n} \cdot \underline{\dot{n}}). \tag{B1}$$

Except the term $I_\parallel \dot{\omega}_\parallel^I$, all other terms in (B1) vanish. Indeed, $\underline{\omega}_\perp^I \cdot \underline{n} = 0$ due to Eq.(1), $\underline{\omega}_\perp^I \cdot \underline{\dot{n}} = \underline{\omega}_\perp^I \cdot (\underline{\omega}_\perp^I \times \underline{n}) = 0$, and $\underline{n} \cdot \underline{\dot{n}} = 0$ since $\underline{n}$ is the unit vector. It proves the first equation in (18).

2) Using Eq.(17) yields:

$$\underline{\underline{\dot{S}}} \cdot \underline{n} = I_\perp[d/dt(\underline{\underline{\omega}}_\perp^I \cdot \underline{n}) - \underline{\underline{\omega}}_\perp^I \cdot \underline{\dot{n}}] + I_\parallel \underline{\underline{\dot{\omega}}}_\parallel^I \cdot \underline{n}. \tag{B2}$$

Due to Eq.(7), $\underline{\omega}_\perp^I \cdot \underline{n} = -\dot{\underline{n}}$ and $\underline{\underline{\omega}}_\perp^I \cdot \dot{\underline{n}} = |\dot{\underline{n}}|^2 \underline{n}$. The last term in Eq.(B2) is calculated as follows: $\underline{\underline{\dot{\omega}}}_\parallel^I \cdot \underline{n} = \dot{\omega}_\parallel^I \underline{n} \times \underline{n} + \omega_\parallel^I \underline{n} \times \dot{\underline{n}} = \omega_\parallel^I \underline{\omega}_\perp^I$. Then Eq.(B2) becomes

$$\underline{\underline{\dot{S}}} \cdot \underline{n} = -I_\perp [\ddot{\underline{n}} + \underline{n} |\dot{\underline{n}}|^2] + I_\parallel \omega_\parallel^I \underline{\omega}_\perp^I \tag{B3}$$

Now we use the condition $n_i n_i = 1$. It leads to relation $|\dot{\underline{n}}|^2 = -\underline{n} \cdot \ddot{\underline{n}}$. Substituting this result into (B3) proves the second equation in (18).

**Appendix C**. Derivation of Eq.(21).

To prove Eq.(21), the rotational kinetic energy is represented in the form:

$$K_r = 1/2 I_\perp (\omega_\perp^I)^2 + 1/2 I_\parallel (\omega_\parallel^I)^2 \tag{C1}$$

Then its time derivative is:

$$\dot{K}_r = I_\perp \underline{\omega}_\perp^I \cdot \underline{\dot{\omega}}_\perp^I + I_\parallel \underline{\omega}_\parallel^I \cdot \underline{\dot{\omega}}_\parallel^I \equiv \underline{\omega}^I \cdot \underline{\dot{L}} - A . \tag{C2}$$

Here $\underline{L}$ is defined in Eq. (17), $\underline{\omega}^I = \underline{\omega}_\perp^I + \underline{\omega}_\parallel^I$, and

$$A \equiv I_\perp \underline{\omega}_\parallel^I \cdot \underline{\dot{\omega}}_\perp^I + I_\parallel \underline{\dot{\omega}}_\parallel^I \cdot \underline{\omega}_\perp^I = (I_\parallel - I_\perp) \underline{\dot{\omega}}_\parallel^I \cdot \underline{\omega}_\perp^I . \tag{C3}$$

In transformation of (C3) we used Eq.(1). The scalar product in (C3), calculated as:

$$\underline{\dot{\omega}}_\parallel^I \cdot \underline{\omega}_\perp^I = d/dt(\omega_\parallel^I \underline{n}) \cdot \underline{\omega}_\perp^I = \dot{\omega}_\parallel^I \underline{n} \cdot \underline{\omega}_\perp^I + \omega_\parallel^I \dot{\underline{n}} \cdot \underline{\omega}_\perp^I = \omega_\parallel^I (\underline{\omega}_\perp^I \times \underline{n}) \cdot \underline{\omega}_\perp^I = 0 ,$$

shows that in Eq. (C2), the term $A = 0$. This proves Eq.(21).